\documentclass[aps,prb,twocolumn,showpacks]{revtex4}
\usepackage{amsmath}
\usepackage{amsfonts}

\newcommand{\s}{{\bf S}}

\newcommand{\ket}[1]{\left|#1\right\rangle}

\newcommand{\bra}[1]{\left\langle#1\right|}
\newcommand{\ketcol}[2]{\ket{\begin{subarray}{c}#1\\[4pt]#2\end{subarray}}}

\begin{document}

\title{Antiferromagnetic ordering of energy levels for spin ladder with four-spin cyclic exchange:
Generalization of the Lieb-Mattis theorem}

\author{Tigran Hakobyan}

\email{hakob@yerphi.am}

\affiliation{%
Yerevan State University, 1 Alex Manoogian, 375025 Yerevan,
Armenia,\\
Yerevan Physics Institute, 2 Alikhanyan Br., 375036 Yerevan, Armenia
}%

\begin{abstract}
The Lieb-Mattis  theorem is generalized to an antiferromagnetic
spin-ladder model with four-spin cyclic exchange interaction. We
prove that for $J > 2 K$, the antiferromagnetic ordering of energy
levels takes place separately in two sectors, which remain symmetric
and antisymmetric under the reflection with respect to the
longitudinal axis of the ladder. We prove also that at the self-dual
point $J = 2 K$, the Lieb-Mattis rule holds in the sectors with
fixed number of rung singlets. In both cases, it agrees with the
similar rule for Haldane chain with appropriate spin number.

\end{abstract}


\date{February 17, 2008}

\maketitle

The systems with multi-spin exchanges have gained a lot of interest
for a long time (for a recent review, see Refs. \onlinecite{Rog05,
Misg03}). These interactions arise at higher orders of a strong
coupling expansion of a half-filled Hubbard model and provide
perturbative corrections to the Heisenberg model. \cite{Tak77} The
multi-spin cyclic permutations first were suggested to be important
in two-dimensional (2D) quantum solids like ${}^3$He. \cite{Th65}
The relevance of the four-spin cyclic interaction around square
plaquettes of CuO${}_2$ planes in high-temperature superconductors
was suggested in Ref. \onlinecite{RD89}, and then it was proven
experimentally. \cite{Rog05} Recently, their quasi-1D counterparts
with similar structure and properties have been studied intensively
(see Ref. \onlinecite{Dag99} for a review). Note that the
spin-ladder model is the simplest system, where the four-spin cyclic
exchange appears from the electron interaction. In fact, the
investigations of some copper-based spin-ladder materials have
revealed the relevance of four-spin exchange term
 \cite{Breh99} (see Ref. \onlinecite{SU05} for a review).
The ground-state phase diagrams of frustrated spin-1/2 ladder
systems are well investigated. \cite{Ners00} The inclusion of
four-spin interactions may result in new unconventional phases.
\cite{NT97,Mom03,Hik03,LT06} The different phases are related by a
duality transformation. \cite{Mom03}

In this paper, we will generalize the well-known Lieb-Mattis
theorem on ordering of energy levels to the ladder model with
four-spin cyclic interaction. For finite-size Heisenberg models on
bipartite lattices, Lieb and Mattis proved that the lowest energy
$E(S)$ in the sector, where the total spin is equal to $S$, is a
monotone increasing function of the spin for any $S\ge
S_{\text{gs}}$, \cite{LM62} where $ S_{\text{gs}}$ is the spin
 of the ground state. This property is known as Lieb-Mattis
theorem about antiferromagnetic ordering of energy levels. The
bipartiteness means that the lattice can be divided into two
sublattices $A$ and $B$, so that all interactions within the same
sublattice are ferromagnetic while the interactions between
different sublattices are antiferromagnetic. Moreover, the quantum
ground state of finite-size system is a unique multiplet with total
spin $S_{\text{gs}}=|S_A-S_B|$, which coincides with the spin of the
classical ground state, namely, the N\'eel state. Here, $S_A$ and
$S_B$ are the highest possible spins on corresponding sublattices.
\cite{LM62}  In one dimension, the Lieb-Mattis theorem is valid for
a more general class of quantum systems. In particular, it is true
for the Hubbard chain. \cite{LM62'} Recently, it has been
generalized to SU($n$) symmetric chain. \cite{H04} A ferromagnetic
ordering of energy levels has also been formulated and proven for
the Heisenberg chain. \cite{Nach03}

According to numerical simulations, a weak frustration may
preserve the antiferromagnetic ordering of energy levels
\cite{Rich95} and the ground-state spin value, \cite{Liu02}
whereas a stronger frustration can violate the Lieb-Mattis
property. For many frustrated systems, the lowest levels $E(S)$
show  approximately parabolic or linear growth. \cite{Schnack01}
So, although the Lieb-Mattis theorem is not valid for frustrated
spin systems in general, it (or its proper extensions) may be
valid for certain systems. In particular, its generalization to
reflection-symmetric frustrated spin-1/2 ladder model has been
formulated and proven recently. \cite{H07} Here, we obtain similar
results for the frustration caused by four-spin ring interaction.

 The Hamiltonian of the system reads:
\begin{equation}
\label{H}
\begin{split}
H&=\sum_{l=1}^{N-1} J^\parallel_l(\s_{1, l}\cdot\s_{1, l+1}+\s_{2, l}\cdot\s_{2, l+1})
\\
&+ \sum_{l=1}^{N} J^\perp_l\s_{1,l}\cdot\s_{2,l}
+ \sum_{l=1}^{N-1}K_l (P_{l,l+1}^\square + P_{l,l+1}^{\square-1}),
\end{split}
\end{equation}
where $\s_{1,l}$ and $\s_{2,l}$ are the spin-1/2 operators of the
first and second chains respectively. The cyclic ring exchange
$P^\square+P^{\square-1}$ is a superposition of clockwise and
counter clockwise permutations of four spins around each
plaquette. We consider the following range of couplings:
\begin{equation}
\label{range}
J^\parallel_l > 2K_l > 0.
\end{equation}

The system possesses SU(2) spin symmetry. It has also $Z_2$
symmetry corresponding to the reflection with respect to the
longitudinal axis. So, the Hamiltonian remains invariant on
individual sectors with fixed values of spin $S$ and reflection
$\sigma=\pm1$ quantum numbers.

We will prove that for the model \eqref{H}, \eqref{range}, the
antiferromagnetic ordering of energy levels holds independently in
symmetric ($\sigma=1$) and antisymmetric ($\sigma=-1$) sectors and
conforms to the similar rule for Haldane chain, \cite{Hal83} i.e.
spin-1 Heisenberg chain, with $N$ and $N-1$ spins, respectively.
Namely, the lowest-energy levels $E_\sigma(S)$ in sectors with spin
$S$ and parity $\sigma$ are nondegenerate and monotone increasing
functions of $S$ for $S\ge S_\text{gs}(\sigma)$. Here,
$$
S_\text{gs}(\sigma)=
\begin{cases}
0, & \text{if $\sigma=(-1)^N$}\\
1, & \text{if $\sigma=(-1)^{N-1}$}\\
\end{cases}
$$
is the ground-state spin value in the sector with parity $\sigma$.
The nondegeneracy  means that all states on corresponding level
form a unique multiplet. So, the ground state in $\sigma=(-1)^N$
sector is a unique singlet  while in $\sigma=(-1)^{N-1}$  sector
it is a unique triplet.

We will prove also that at the self-dual point $J^\parallel_l =
2K_l$, the Lieb-Mattis rule holds in the sectors with fixed number
of rung singlets $N_0$ and agrees with the similar rule for the
Haldane chain with $N-N_0$ spins.

%

We begin by introducing the basis of three triplet and one singlet
states for each rung:
\begin{equation}
\begin{split}
\label{rung}
&\ket{1}=\ketcol{\uparrow}{\uparrow}, \quad
\ket{\tilde{0}}=\frac1{\sqrt{2}}
\left(\ketcol{\uparrow}{\downarrow}+\ketcol{\downarrow}{\uparrow}\right),
\quad \ket{-1}=\ketcol{\downarrow}{\downarrow},
\\
&\ket{0}=\frac1{\sqrt{2}}\left(\ketcol{\uparrow}{\downarrow}-\ketcol{\downarrow}{\uparrow}\right).
\end{split}
\end{equation}
Below, we will prove that all nonvanishing off-diagonal elements
of the Hamiltonian \eqref{H},\eqref{range} become negative in the
basis
\begin{equation}
\begin{split}
\label{neg-basis}
&\ket{m_1,\ldots,m_{N}}
\\
&\qquad=(-1)^{[N_0/2]+N_{0\tilde0}+M_\text{odd}} \ket{m_1}\otimes\ldots\otimes\ket{m_{N}},
\end{split}
\end{equation}
where $ m_l=\pm1, \tilde0,0$ marks one of the four rung states
\eqref{rung}. In the sing factor, $N_0$ is the number of singlets,
and  $N_{0\tilde0}$ is the number of pairs $(0,\tilde0)$ in the
sequence $m_1,\dots,m_N$, where $0$ is on the left-hand side from
$\tilde 0$. \cite{H07} $M_\text{odd}=\sum_l m_{2l-1}$ is the total
$z$-projection of odd-site spins. Note that the basic states above
are eigenstates of the reflection operator with eigenvalue
$\sigma=(-1)^{N_0}$.

First, we rewrite the Hamiltonian in a form that is more convenient
for further purposes. The four-spin interaction term can be
expressed via spin operators as follows (see, for instance, Ref.
\onlinecite{GNB04}):
\begin{equation*}
\begin{split}
P_{l,l+1}^\square &+ P_{l,l+1}^{\square-1} = \s_{1,l}\cdot\s_{1,l+1}
 +\s_{2,l}\cdot\s_{2,l+1}  +\s_{1,l}\cdot\s_{2,l}
  \\
&+  \s_{1,l+1}\cdot\s_{2,l+1} + \s_{1,l}\cdot\s_{2,l+1}
+\s_{1,l+1}\cdot\s_{2,l}
\\
& + 4(\s_{1,l}\cdot\s_{1,l+1})(\s_{2,l}\cdot\s_{2,l+1})
\\
& + 4(\s_{1,l}\cdot\s_{2,l})(\s_{1,l+1}\cdot\s_{2,l+1})
\\
& - 4(\s_{1,l}\cdot\s_{2,l+1})(\s_{2,l}\cdot\s_{1,l+1})+1/4.
\end{split}
\end{equation*}
Using the relation $\s_1\cdot\s_2=P_{12}/2-1/4$, where the
operator $P_{12}$ permutes two spin states, one can present the
expression above in the following form:
\begin{equation}
\begin{split}
\label{cyclic} P_{l,l+1}^\square + P_{l,l+1}^{\square-1} &=
2(\s_{1,l}\cdot\s_{2,l+1}+\s_{1,l+1}\cdot\s_{2,l})
\\
&+ P^{\, \parallel}_{l,l+1}+P^{\, =}_{l,l+1}-P^\times_{l,l+1}.
\end{split}
\end{equation}
Here, $P^{\, \parallel}_{l,l+1}$, $P^{\, =}_{l,l+1}$ and
$P^\times_{l,l+1}$ are, respectively, four-spin permutations along
the plaquette rungs, legs and diagonals.
%
Further, we express the two-spin interactions in terms of the
symmetrized and antisymmetrized rung spin operators
\begin{equation}
\label{S-sa} \s^{(s)}_l=\s_{1,l}+\s_{2,l}, \qquad
\s^{(a)}_l=\s_{1,l}-\s_{2,l}.
\end{equation}
The operator $\s^{(s)}_l$ describes the total spin of $l$th rung.
Using \eqref{cyclic}, \eqref{S-sa} and omitting nonessential
scalar term, one can reduce the Hamiltonian \eqref{H} to the
following form:
\begin{equation}
\begin{split}
\label{H-s-a} H &= \sum_{l=1}^{N-1}(J^s_l\,
\s^{(s)}_l\cdot\s^{(s)}_{l+1}+J^a_l\,
\s^{(a)}_l\cdot\s^{(a)}_{l+1})
\\
&+ \sum_{l=1}^{N-1} K_l(P^{\, \parallel}_{l,l+1}+P^{\,
 =}_{l,l+1}-P^\times_{l,l+1})
\\
&+ \sum_{l=1}^{N} \frac{J^\perp_l}{2}(\s^{(s)}_l)^2.
\end{split}
\end{equation}
In the above equation, we have introduced the symmetrized and
antisymmetrized couplings
\begin{equation}
\label{Jsa} J^s_l=\frac{J^\parallel_l}2+K_l, \qquad
J^a_l=\frac{J^\parallel_l}2-K_l.
\end{equation}
Note that a similar decomposition for the Hamiltonian without
four-spin exchange was applied in Refs. \onlinecite{Xian95} and
\onlinecite{LT06}. The permutations $P^{\, \parallel}_{l,l+1}$ and
$P^{\, \times}_{l,l+1}$ have been used in Ref. \onlinecite{GNB04}.

The $J^\perp$ part of the Hamiltonian is just the sum of rung spins
squares, which is diagonal in the basis \eqref{neg-basis}. The local
terms $P^{\,\parallel}_{l,l+1}$ are also diagonal (with eigenvalues
$\pm1$), since any triplet (singlet) rung state stays symmetric
(antisymmetric) under the reflection.

The $J^s$ terms correspond to the so-called composite spin model.
\cite{ST86} They conserve the spins of individual rungs because
$\s^{(s)}_l$ describes the total spin of $l$th rung. \cite{Xian95}
The singlets remain frozen at their points, and, therefore, the
factor $(-1)^{[N_0/2]+N_{0\tilde0}}$ in \eqref{neg-basis} remains
invariant. All nonvanishing off-diagonal matrix elements come from
the exchanges
$\ket{\tilde0}\otimes\ket{\tilde0}\leftrightarrow\ket{\pm1}\otimes\ket{\mp1}$
and
$\ket{\tilde0}\otimes\ket{\pm1}\leftrightarrow\ket{\pm1}\otimes\ket{\tilde0}$
of two neighboring triplet states, which alter the sign of
$(-1)^{M_\text{odd}}$. Note that they coincide with similar matrix
elements of the Haldane chain, in (nonpositive) basis formed by the
states
$(-1)^{M_\text{odd}}\ket{m_1}\otimes\ldots\otimes\ket{m_{N}}$.
\cite{LM62} So, the composite spin part of the Hamiltonian is
nonpositive in the basis \eqref{neg-basis}.

The matrix elements produced by the antisymmetric local terms of
Hamiltonian \eqref{H-s-a} have been considered already in Ref.
\onlinecite{H07}. In terms of lowering-raising operators
$S^{(a)\pm}=S^{(a)x}\pm iS^{(a)y}$, each such term reads
$(S^{(a)+}_lS^{(a)-}_{l+1}+S^{(a)-}_lS^{(a)+}_{l+1})/2+S^{(a)z}_lS^{(a)z}_{l+1}$.
In contrary to the symmetric case, the antisymmetrized spin
operators mix triplet and singlet states. Their nonzero matrix
elements are: \cite{Noak06}
\begin{equation}
\begin{split}
\label{s-asym}
&\bra{0}S^{(a)+}\ket{-1}=\sqrt{2}, \quad
\bra{1}S^{(a)+}\ket{0}=-\sqrt{2},
\\
&\bra{\tilde0}S^{(a)z}\ket{0}=1.
\end{split}
\end{equation}
Using the above equations and the definition of basic states
\eqref{neg-basis}, it is easy to check that
\begin{equation}
\label{s-pm-a}
\begin{split}
\bra{\dots,0_l,0_{l+1},\dots}S^{(a)\mp}_l S^{(a)\pm}_{l+1} \ket{\dots,\pm1_l,\mp1_{l+1},\dots}
\\
=-2(-1)^{[N_0/2]+N_{0\tilde0}+M_\text{odd}-[N'_0/2]-N'_{0\tilde0}-M'_\text{odd}}=-2,
\end{split}
\end{equation}
where the unchanged sites are replaced by dots. The quantum numbers
of bra- and -ket states are mentioned, respectively, without and
with primes. Indeed, according to the definition of $N_{0\tilde0}$,
the difference $N_{0\tilde0}-N'_{0\tilde0}$ in \eqref{s-pm-a} is an
even number. Also, $N'_0=N_0+2$ and $M_\text{odd}=M'_\text{odd}\pm1$
depending on whether $l$ is even or odd. Therefore, the exponent in
\eqref{s-pm-a} is an even number, and the equation is true. The next
nontrivial matrix element is also negative. Namely,
\begin{equation}
\label{s-pm-b}
\begin{split}
\bra{\dots,0_l,\pm1_{l+1},\dots}S^{(a)\mp}_l S^{(a)\pm}_{l+1} \ket{\dots,\pm1_l,0_{l+1},\dots}
\\
=2(-1)^{M_\text{odd}-M'_\text{odd}}=2(-1)^{\pm1}=-2
\end{split}
\end{equation}
because the quantum numbers $N_{0\tilde0}$ and $N_0$ are the same for both states.

In contrast, the $z$ projections of antisymmetrized spin operators
preserve the quantum number $M_\text{odd}$. They produce negative
matrix elements too:
\begin{equation}
\label{s-z-a}
\begin{split}
\bra{\dots,0_l,0_{l+1},\dots}S^{(a)z}_l S^{(a)z}_{l+1} \ket{\dots,\tilde0_l,\tilde0_{l+1},\dots}
\\
=(-1)^{N_{0\tilde0}-N'_{0\tilde0}+(N_0-N'_0)/2}=(-1)^{\text{even}-1}=-1
\end{split}
\end{equation}
and
\begin{equation}
\label{s-z-b}
\begin{split}
\bra{\dots,\tilde0_l,0_{l+1},\dots}S^{(a)z}_l S^{(a)z}_{l+1} \ket{\dots,0_l,\tilde0_{l+1},\dots}
\\
=(-1)^{N_{0\tilde0}-N'_{0\tilde0}}=(-1)^1=-1.
\end{split}
\end{equation}
The expressions \eqref{s-pm-a},  \eqref{s-pm-b}, \eqref{s-z-a},
\eqref{s-z-b} together with conjugate ones constitute the full set
of nontrivial matrix elements of Hamiltonian \eqref{H-s-a}
generated by $J^a$ terms.

Finally, consider the off-diagonal terms, which are responsible for
four-spin cyclic exchange. The operator $P^{\,=}_{l,l+1}$ just
permutes two adjacent rung states. At the same time,
$P^\times_{l,l+1}$ is a signed permutation: While permuting the
singlet with a triplet state it produces an additional minus sign.
Therefore, the difference $P^{\,=}_{l,l+1}-P^\times_{l,l+1}$
vanishes if the spins of both rungs are the same. If their spins
differ, this operator just permutes them multiplying by $2$. Then,
using the definition \eqref{neg-basis} of basic states, we obtain
\begin{equation}
\label{k-term}
\begin{split}
\bra{\dots,t_l,0_{l+1},\dots} P^{\,=}_{l,l+1}-P^\times_{l,l+1} \ket{\dots,0_l,t_{l+1},\dots}
\\
=2(-1)^{M_\text{odd}-M'_\text{odd}+N_{0\tilde0}-N'_{0\tilde0}}=-2,
\end{split}
\end{equation}
where $t=\tilde0,\pm1$ is any triplet state. Indeed, in the sign
factor above, $M_\text{odd}=M'_\text{odd}$ and
$N_{0\tilde0}-N'_{0\tilde0}=1$ for $t=\tilde0$. For $t=\pm1$,
$N_{0\tilde0}=N'_{0\tilde0}$ and $|M_\text{odd}-M'_\text{odd}|=1$.
Together with the conjugate element, this is a sole nonvanishing
off-diagonal matrix element produced by the four-spin cyclic
exchange term.

According to the constraints \eqref{range} imposed on the couplings,
the coefficients $J^s_l,J^a_l,K_l$ in \eqref{H-s-a} are positive.
This finishes the proof that the ladder Hamiltonian has no positive
off-diagonal element in the basis \eqref{neg-basis}.

Due to the spin and reflection symmetries, the Hamiltonian is
invariant on each $(M,\sigma)$ subspace, all states of which have
$S^z=M$ and $\sigma=\pm1$ quantum numbers. Any two basic states
within the same subspace are connected at least by two-spin
interaction terms of the Hamiltonian, as can be easily verified by
induction. \cite{H07} So, we can apply the Perron-Frobenius theorem
\cite{PF} to the matrix of the Hamiltonian restricted to any
$(M,\sigma)$ subspace. As a result, the lowest energy state there
(usually called a relative ground state) is unique and is a positive
superposition of all basic states:
\begin{equation}
\label{gs} \ket{\Omega}_{M,\sigma}=
\sum_{\ket{m_1,\dots,m_{N}}\in (M,\sigma)}
\omega_{m_1\dots m_N}\ket{m_1,\dots,m_{N}},
\end{equation}
where $\omega_{m_1\dots m_N}>0$. The uniqueness implies that this
state must have a certain value of spin $S_{M,\sigma}$, which can be
obtained by comparing with the similar state of the Haldain chain.
The last model corresponds to the restriction of the composite spin
model $\sum_l \s^{(s)}_l\cdot\s^{(s)}_{l+1}$ on the states with
triplets on all rungs. In fact, all such type states from
\eqref{neg-basis} form a nonpositive basis for the Haldane chain.
\cite{LM62} Its relative ground state $\ket{\Omega}_M$ in $S^z=M$
subspace is a positive superposition of all basic states and has the
highest possible spin value, i.e., $|M|$, except $M=0$ and odd $N$
case when its spin is one. \cite{LM62} Therefore, for $\sigma=1$,
both states $\ket{\Omega}_{M,\sigma}$ and $\ket{\Omega}_M$ overlap,
and, hence, must have the same spin. Similarly, the restriction of
composite spin model to the subspace of states with one singlet
frozen at the last rung corresponds to the Haldane chain with $N-1$
sites. So, for $\sigma=-1$, the spin of \eqref{gs} coincides with
the spin of the corresponding state of the Haldane chain having one
site less. Therefore, the spin of the relative ground state
\eqref{gs} is $|M|$ except $M=0$ and $\sigma=(-1)^{N-1}$ case when
it equals one.

Now we are ready to finish the proof of our main result.
 For $\sigma=(-1)^N$, the relative ground states
$\ket{\Omega}_{\pm S,\sigma}$ are, correspondingly, the highest and
lowest states of a unique spin-$S$  multiplet, which has the minimum
energy $E_\sigma(S)$ among all spin-$S$ states with parity $\sigma$.
The $(S,\sigma)$ subspace contains a representative from any
multiplet with $S'\ge S$ and parity $\sigma$. Together with the
uniqueness condition, this implies that $E_\sigma(S')$ must be
higher than $E_\sigma(S)$ for all $S'>S$. Consequently,
$E_\sigma(S)$ is a monotone increasing function of $S$, and the
ground state in $\sigma=(-1)^N$ sector is a nondegenerate spin
singlet. For $\sigma=(-1)^{N-1}$, the states $\ket{\Omega}_{\pm
S,\sigma}$ are the highest and lowest states only if $S\ge 1$.
Therefore, $E_\sigma(S)$ monotonically increases in this region.
Note that the states $\ket{\Omega}_{\pm 1,\sigma}$ and
$\ket{\Omega}_{0,\sigma}$, which have the lowest energy in
$\sigma=(-1)^{N-1}$ sector, form a spin triplet.

Consider separately the limiting case of $J^\parallel_l=2K_l$ when
the $J^{a}$ terms in \eqref{H-s-a} are absent. Then $\sum_l
(\s^{(s)}_l)^2$ commutes with the Hamiltonian and with the total
spin operator. \cite{Mom03} As a result, the symmetry group
SU(2)$\times Z_2$ expands to SU(2)$\times$U(1)=U(2). The U(1)
symmetry reflects the invariance under the duality transformation
and results in the conservation of singlet number $N_0$.
\cite{Hik03} Any $(M,\sigma)$ subspace splits into invariant
subspaces with fixed singlet number obeying $(-1)^{N_0}=\sigma$. It
is easy to see that the Hamiltonian is connected on every such
subspace. Therefore, the relative ground state there is unique and
is a positive superposition of all basic states \eqref{neg-basis}
with $S^z=M$ and $N_0$ rung singlets. Again, comparing it with the
action of the composite spin model on the states with singlets on
the last $N_0$ rungs, one can conclude that the relative ground
state has the highest possible spin value, except $M=0$ and odd
$N-N_0$ case when it is a triplet state. The antiferromagnetic
ordering of energy levels takes place independently in any sector
with fixed singlet number and corresponds to the similar rule for
the Haldane chain with $N-N_0$ spins.

For appropriate values of couplings, the results of this paper
remain true, if diagonal interactions
$\s_{1,l}\cdot\s_{2,l+1}+\s_{1,l+1}\cdot\s_{2,l}$ are included in
the Hamiltonian \eqref{H}. Similarly, one can consider a more
general biquadratic interaction of type $
K'_l(\s_{1,l}\cdot\s_{1,l+1})(\s_{2,l}\cdot\s_{2,l+1})
 + 4K_l(\s_{1,l}\cdot\s_{2,l})(\s_{1,l+1}\cdot\s_{2,l+1})
- 4K_l(\s_{1,l}\cdot\s_{2,l+1})(\s_{2,l}\cdot\s_{1,l+1})
$
with arbitrary couplings  $K'_l$ because the first term
in the sum is diagonal in the basis \eqref{rung}, \eqref{neg-basis}.

\begin{acknowledgments}
The author is grateful to V.~Ohanyan for useful discussions.   This
work was supported by grants UC-06/07, INTAS-05-7928, ANSEF-1386-PS
and the Artsakh Ministry of Science \& Education.
\end{acknowledgments}

\end{document}